\documentclass[aps,pra,twocolumn,showpacs,floatfix]{revtex4}

\usepackage{graphicx}
\usepackage{nicefrac}
\usepackage{amsmath}
\usepackage{amsfonts}
\usepackage{amssymb}
\usepackage{amsthm}
\usepackage{epsf}
\usepackage{bm}
\usepackage{bbm}
\usepackage{longtable}

\usepackage{dcolumn}

\def\Za{{Z\alpha}}
\def\pr{^{\prime}}
\def\rms{\lbr r^2 \rbr^{1/2}}
\def\rmsSq{\lbr r^2 \rbr}

\newcolumntype{.}{D{x}{}{-1}}

\newcommand{\vare}{\varepsilon}
\newcommand{\bfr}{{\bm {r}}}

\newcommand{\bfp}{{\bm {p}}}
\newcommand{\bfq}{{\bm {q}}}

\newcommand{\hr}{\hat{\bfr}}

\newcommand{\lbr}{\langle}
\newcommand{\rbr}{\rangle}

\newcommand{\balpha}{\bm{\alpha}}
\newcommand{\bsigma}{\bm{\sigma}}
\newcommand{\bmu}{\bm{\mu}}

\newcommand{\SixJ}[6]{
        \left\{
        \begin{array}{ccc}
        #1  & #2  & #3 \\
        #4  & #5  & #6 \\
        \end{array}
        \right\}
        }

\begin{document}

\title{Hyperfine structure of Li and Be$^{\bm{+}}$}

\author{V.~A.~Yerokhin}
 \affiliation{Center for Advanced Studies, St.~Petersburg State Polytechnical
University, Polytekhnicheskaya 29, St.~Petersburg 195251, Russia}

\begin{abstract}

A large-scale relativistic configuration-interaction (CI) calculation is performed
for the magnetic-dipole and the electric-quadrupole hyperfine structure splitting
in $^{7,6}$Li and $^9$Be$^+$. Numerical results for the $2^2S$, $3^2S$,
$2^2P_{1/2}$, and $2^2P_{3/2}$ states are reported. The CI calculation based on
the Dirac-Coulomb-Breit Hamiltonian is supplemented with separate treatments of
the QED, nuclear-magnetization distribution, recoil, and negative-continuum
effects.

\end{abstract}


\maketitle

\section{Introduction}

The hyperfine structure (hfs) of few-electron atoms has been an attractive subject
of theoretical studies for decades, one of the reasons being a few ppm accuracy
achieved in experiments on Li and Be$^+$ \cite{beckmann:74,wineland:83:prl}.
Despite the considerable attention received, a high-precision theoretical
determination of hfs in few-electron atoms remaines a difficult task. The main
problem lies in the high singularity of the hfs interaction and, as a consequence,
in the dependence of the calculated results on the quality of the many-electron
wave function near the nucleus.

The hfs splitting of lithium has traditionally been one of the standard test cases
for different theoretical methods \cite{hibbert:75}. Among various calculations
reported in the literature, the nonrelativistic ones are the most numerous; their
technique is well developed by now. The best numerical accuracy for the
nonrelativistic hfs value is achieved in variational calculations that use
multiple basis sets in Hylleraas coordinates \cite{king:89,yan:96,king:07}.
Probably the most popular nonrelativistic approach is the multiconfigurational
Hartree-Fock (MCHF) method \cite{carlsson:92:pra,godefroid:01}, which is less
computationally intensive but also produces less accurate results. The main
drawback of the nonrelativistic methods is that the relativistic effects should be
accounted for separately. There is a way to perform a direct evaluation of the
leading relativistic correction \cite{pachucki:02:pra}, but such a calculation is
difficult and has not yet been done. So far, the relativistic correction was
estimated by comparing with less accurate calculations based on the
Dirac-Coulomb-Breit Hamiltonian, or by re-scaling the hydrogenic correction.

There were calculations performed for Li and Be$^+$ with the relativistic analog
of the MCHF procedure, the multiconfigurational Dirac-Fock (MCDF) method
\cite{bieron:96,bieron:99,boucard:00}. The computational accuracy of these
relativistic calculations turns out be lower than that of the best nonrelativistic
studies. This is to a large extent due to the fact that the electron correlation
is more difficult to be accurately accounted for relativistically than
nonrelativistically.

Methods that allow a straightforward generalization to the relativistic case are
many-body perturbation theory \cite{lindgren:84:rpp} and its all-order extensions
known as the coupled-cluster (CC) approach
\cite{lindgren:85:pra,blundell:89:pra,martensson-pendrill:90,pal:07,johnson:08}.
Calculations of the Li hfs performed with these methods so far did not account for
the most part of the triple excitations (and, in most cases, for a part of the
double excitations as well), which led to an incomplete treatment of the electron
correlation and to a relatively low accuracy of the corresponding results.
Significant progress in the CC calculations was reported in Ref.~\cite{porsev:06}
for the case of Na. In that work, all valence triple excitations were included.
Such an approach, when applied to the Li hfs, would significantly improve the
accuracy of the CC results. The corresponding calculation is presently underway
\cite{derevianko:priv}.

In the current investigation, a relativistic calculation of the hfs splitting in
Li and Be$^+$  will be performed by employing the configuration-interaction (CI)
method. Unlike the MCDF procedure, the CI method does not involve a variational
minimization and thus is not handicapped by the danger of the variational collapse
into the negative continuum (which manifests itself in a ``sinking'' of the
ground-state energy due to the admixture of the negative energy states into the
ground-state wave function). The CI method has a potential to be more accurate
than the MCDF method. The only problem is that it requires the Dirac spectrum to
be sufficiently well represented by the model space of one-electron wave
functions, whereas the MCDF method can produce reasonable results with only a few
configurations. While this might come as a limitation in the case of complicated
many-electron atoms, the systems at hand, the Li-like atoms, are sufficiently
simple to be very accurately described by the CI method.

The goal of the present investigation is to perform a calculation  of the hfs in
Li and Be$^+$ complete to the relative order $\alpha^2$, where $\alpha$ is the
fine-structure constant. Such a calculation requires, besides a high-precision
determination of the dominant nonrelativistic contribution, a rigorous treatment
of the leading relativistic correction $\sim$$\alpha^2$ and the inclusion of the
QED effects $\sim$$\alpha$ and $\sim$$\alpha^2$. Nuclear effects (the recoil and
the magnetization distribution) also contribute on this level. Because of their
smallness, these effects can be treated nonrelativistically. Since we are
concerned with the effects of order up to $\alpha^2$ only, the Dirac-Coulomb-Breit
Hamiltonian may be used as a convenient and sound starting point for our
investigation.

The calculation complete to the relative order $\alpha^2$ was reported for the hfs
splitting of the $2^2S$ and $3^2S$ states of Li and Be$^+$ in our previous paper
\cite{yerokhin:08:pra}. In this work, we extend our calculations to the $2^2P_J$
states and present a detailed analysis of various corrections. In particular, the
recoil correction to the magnetic dipole hfs interaction is derived for the case
of an arbitrary spin of the nucleus. This correction is shown to yield the
dominant recoil contribution for the hfs splitting of the $P$ and higher-$l$
states in medium-$Z$ H-like atoms. To the best of our knowledge, it has not
previously been accounted for in systems other than hydrogen and deuterium.

The paper is organized as follows. In Sec.~\ref{sec:CI}, a brief summary of the CI
method is given. In Sec.~\ref{sec:hfs}, we present some basic formulas for the
magnetic-dipole and the electric-quadrupole interaction and describe our CI
calculation of the hfs splitting based on the Dirac-Coulomb-Breit Hamiltonian.
Various corrections to the hfs are calculated in Sec.~\ref{sec:corrections}. The
results obtained are discussed and compared with the experimental data in
Sec.~\ref{sec:discussion}.

Relativistic units $\hbar = c = 1$ and $\alpha = e^2/(4\pi)$ are used throughout
this paper.


\section{Configuration interaction method}
\label{sec:CI}

Relativistic Hamiltonian of an $N$-electron atom can be written as
\begin{equation}\label{eq1}
    H_{DCB} = \sum_i h_{\rm D}(i) + \sum_{i<j} \left[ V_{C}(i,j)+
    V_{B}(i,j)\right]\,,
\end{equation}
where indices $i,j = 1,\ldots,N$ numerate the electrons, $h_D$ is the one-particle
Dirac Hamiltonian,
\begin{equation}\label{eq2}
    h_D(i) = \balpha_i\cdot\bfp_i+ (\beta-1)\,m+ V_{\rm nuc}(r_i)\,,
\end{equation}
$\balpha$ and $\beta$ are the Dirac matrices, $V_{\rm nuc}$ is the binding
potential of the nucleus, $V_{C}(i,j) = \alpha/r_{ij}$ is the Coulomb part of the
electron-electron interaction, $r_{ij} = |\bm{r}_i-\bm{r}_j|$, $V_{B}$ is the
Breit interaction,
\begin{equation}\label{eq3}
V_{B}(i,j) =  -\frac{\alpha}{2\,r_{ij}}\,
    \left[ \balpha_i\cdot\balpha_j + \left( \balpha_i\cdot \hr_{ij}\right)
             \left( \balpha_j\cdot \hr_{ij}\right) \right]\,,
\end{equation}
and $\hr = \bfr/r$. It is assumed that $H_{DCB}$ acts in the space of the
positive-energy eigenfunctions of some one-particle Hamiltonian.

The $N$-electron wave function of the system with parity $P$, angular momentum
quantum number $J$, and its projection $M$ is represented as a linear combination
of configuration-state functions (CSFs),
\begin{equation}\label{eq4}
  \Psi(PJM) = \sum_r c_r \Phi(\gamma_r PJM)\,,
\end{equation}
where $\gamma_r$ denotes the set of additional quantum numbers that determine the
CSF. The CSFs are constructed as antisymmetrized products of one-electron orbitals
$\psi_n$ of the form
\begin{equation}\label{eq5}
    \psi_n(\bfr) = \frac1r\,         \left(
        \begin{array}{c}
        G_n(r)\,\chi_{\kappa_n m_n}(\hr)\\
      i F_n(r)\,\chi_{-\kappa_n m_n}(\hr)   \\
        \end{array}
        \right)\,,
\end{equation}
where $\chi_{\kappa m}$ the spin-angular spinor \cite{rose:61}, $\kappa =
(-1)^{j+l+1/2}(j+1/2)$ is the relativistic angular parameter, and $m$ is the
angular momentum projection. In the present work, we chose the one-electron
orbitals $\psi_n$ to be the (positive-energy) eigenfunctions of the one-electron
Dirac Hamiltonian with the frozen-core Dirac-Fock potential,
\begin{equation}\label{eq5a}
    h_{DF} = \balpha\cdot\bfp+ (\beta-1)\,m+ V_{\rm nuc}(r)+
    V^{N-1}_{DF}(\bfr)\,.
\end{equation}
The (non-local) potential $V^{N-1}_{DF}$ is defined by its action on a wave
function,
\begin{align}\label{eq5b}
    V^{N-1}_{DF}(\bfr_1)\,\psi(\bfr_1) &\, = \sum_c \int d\bfr_2\,
      \psi_c^+(\bfr_2)\,\frac{\alpha}{r_{12}}\,
 \nonumber \\ &\, \times
        \Bigl[\psi_c(\bfr_2)\,\psi(\bfr_1)-\psi_c(\bfr_1)\,\psi(\bfr_2)\Bigr]\,,
\end{align}
where the index $c$ runs over the core orbitals. The eigenfunctions of the
Hamiltonian $h_{DF}$ form a complete and orthogonal basis of one-electron
orbitals.

In the CI method, the ionization energy of the system and the mixing coefficients
$c_r$ in Eq.~(\ref{eq4}) are obtained by solving the secular equation
\begin{equation}\label{eq6}
    {\rm det} \bigl\{\lbr \gamma_r PJM|H_{DCB}|\gamma_s PJM\rbr -E_r\,\delta_{rs}\bigr\} =
    0\,.
\end{equation}
The matrix elements of the Hamiltonian between the CSFs can be represented as
linear combinations of the one- and two-particle radial integrals,
\begin{align}\label{eq7}
\lbr \gamma_r PJM| &\,  H_{DCB}|\gamma_s PJM\rbr = \sum_{ab}
 d_{rs}(ab)\,I(ab)
  \nonumber \\
 + &\, \alpha \sum_k \sum_{abcd} v_{rs}^{(k)}(abcd)\,
  \left[R_{k}^C(abcd)+R_{k}^B(abcd)\right]\,.
\end{align}
Here, $a$, $b$, $c$, and $d$ specify the one-electron orbitals, $d_{rs}$ and
$v^{(k)}_{rs}$ are the angular coefficients, $I(ab)$ are the one-electron radial
integrals, and $R_{k}^{C}(abcd)$ and $R_{k}^{B}(abcd)$ are the Coulomb and Breit
two-electron radial integrals. The radial integrals are defined by
\begin{equation}\label{eq7a}
    \lbr a|h_D|b\rbr = \delta_{\kappa_a,\kappa_b}\, \delta_{m_a,m_b}\,I(ab)\,,
\end{equation}
\begin{align}\label{eq7b}
    \lbr ab|V_{C,B}|cd\rbr &\, = \alpha \sum_{k m_k}
     \frac{(-1)^{k-m_k+j_c-m_c+j_d-m_d}}{2k+1}\,
  \nonumber \\ & \times
     C^{k\,m_k}_{j_a\,m_a,j_c\,-m_c}
        \,C^{k\,m_k}_{j_d\,m_d, j_b\,-m_b}\,R_{k}^{C,B}(abcd)\,,
\end{align}
where $C^{j\,m}_{j_1\,m_1, j_2\,m_2}$ are the Clebsch-Gordan coefficients. After
integrating over angular variables, the expression for the one-particle integral
reads
\begin{align}\label{eq8}
    I(ab) &\, = \int_0^{\infty}dr\,
    \Bigl[F_a\left(\frac{d}{dr}+\frac{\kappa}{r}\right)G_b
     -G_a\left(\frac{d}{dr}-\frac{\kappa}{r}\right)F_b
 \nonumber \\ &
     + (G_aG_b+F_aF_b)V_{\rm nuc} -2m F_aF_b \Bigr]\,.
\end{align}
The Coulomb integral is given by
\begin{align} \label{eq10}
  R_{k}^C(abcd) =  &\, (-1)^k\lbr \kappa_a||{\bm{C}}^{(k)}||\kappa_c\rbr \, \lbr \kappa_b||{\bm{C}}^{(k)}||\kappa_d\rbr
  \nonumber \\ & \times
          \int_0^{\infty} dr_1\,dr_2\,
          \frac{r_<^k}{r_>^{k+1}}\,W_{ac}(r_1)\,W_{bd}(r_2)\,,
\end{align}
where $W_{ab} = G_aG_b+F_aF_b$ and ${\bm{C}}^{(J)}$ is the spherical tensor with
components $C^{(J)}_{M}(\hr) = \sqrt{4\pi/(2J+1)}\,Y_{JM}(\hr)$. The expression
for the Breit integral is more complex; it can be found in
Ref.~\cite{johnson:88:b}. The angular coefficients $d_{rs}$ and $v_{rs}^{(k)}$ can
be evaluated analytically \cite{grant:73:cpc,grant:76:jpb}. In the general case,
formulas for them are rather cumbersome. A number of packages is available in the
literature for the numerical evaluation of the angular coefficients
\cite{grant:73:cpc,pyper:78:cpc,grant:80:cpc,gaigalas:01:cpc,gaigalas:02:cpc}.


\section{Hyperfine splitting}
\label{sec:hfs}

\subsection{Magnetic dipole hyperfine splitting}

The relativistic Fermi-Breit operator of the magnetic dipole hyperfine interaction
is given by
\begin{equation}\label{2eq1}
    H_{M1} = \frac{|e|}{4\pi}\, \bmu \cdot {\bm T}^{(1)}\,,
\end{equation}
where $\bmu$ is the operator of the nuclear magnetic moment, acting in the nuclear
subspace. The operator ${\bm T}^{(1)}$ acts in the electronic subspace; it is
given by the sum of the one-electron operators ${\bm t}^{(1)}(i)$,
\begin{equation}\label{2eq2}
   {\bm T}^{(1)} = \sum_i {\bm t}^{(1)}(i) = \sum_i \frac{\bfr_i \times
   \balpha_i}{r_i^3}\,.
\end{equation}
In the nonrelativistic limit, the operator ${\bm t}^{(1)}$ turns into ${\bm
t}^{(1)}_{NR}$,
\begin{equation}\label{2eq2a}
    {\bm t}^{(1)}_{NR} =  \frac1m \left[\frac{\bm l}{r^3}+
    \frac{3\,\hr ({\bf s}\cdot \hr)-{\bm s}}{r^3}+
    \frac{8\pi}{3}\,\delta(\bfr)\,{\bm s} \right]\,,
\end{equation}
where $\bm l$ and $\bm s$ are the one-electron operators of orbital angular
momentum and spin, respectively. The three terms in the brackets are often
referred to as the orbital, the spin-dipole, and the Fermi contact term,
respectively:
\begin{equation}\label{2eq2b}
   {\bm t}^{(1)}_{NR} = {\bm t}^{(1)}_{l}+ {\bm t}^{(1)}_{sd}+  {\bm
   t}^{(1)}_{c}\,.
\end{equation}

The relativistic value of the energy shift due to the magnetic dipole hyperfine
interaction is obtained as the expectation value of the Fermi-Breit operator on
the wave function of the system with atomic angular momentum $F$ and its
projection $M_F$. Employing the Wigner-Eckart theorem, the nuclear variables are
separated and integrated out and the energy shift is represented in terms of the
reduced matrix element of the operator ${\bm T}^{(1)}$,
\begin{align}\label{2eq3}
    \Delta E_{M1} &\, = \lbr FM_F|H_{M1}|FM_F\rbr =
        \frac{|e|}{4\pi}\, \frac{\mu}{2I}\,
 \nonumber \\ & \times
        [F(F+1)-I(I+1)-J(J+1)]\,
 \nonumber \\ & \times
        \frac{ \lbr J||{\bm T}^{(1)}||J\rbr}{\sqrt{J(J+1)(2J+1)}}\,,
\end{align}
where $\mu$ is the magnetic moment of the nucleus, $\mu=\lbr
II|\bm{\mu}_0|II\rbr$, $I$ is the nuclear spin, and $J$ is the total angular
momentum of the electron. Experimental data for the magnetic dipole hfs splitting
are usually expressed in terms of the hyperfine interaction constant $A_J$, which
does not depend on $F$,
\begin{equation}\label{2eq3a}
    A_J =
     \frac{\Delta E_{M1}}{\lbr FM_F|{\bm I}\cdot {\bm J}|FM_F\rbr }
    = \frac{|e|}{4\pi}\, \frac{\mu}{I}\,
        \frac{ \lbr J||{\bm T}^{(1)}||J\rbr}{\sqrt{J(J+1)(2J+1)}}\,.
\end{equation}

The reduced matrix element of the operator ${\bm T}^{(1)}$ should be evaluated
with the CI many-electron wave functions (\ref{eq4}), obtained by solving the
secular equation (\ref{eq6}). Matrix elements of the operator ${\bm T}^{(1)}$
between individual CSFs can be expressed as linear combinations of matrix elements
of the one-electron operator ${\bm t}^{(1)}$ between the single-particle orbitals,
\begin{equation}\label{2eq4}
  \lbr \gamma_r PJ||{\bm T}^{(1)}||\gamma_s PJ\rbr = \sum_{a \le b}
  \,d_{rs}^{(1)}(ab)\, \lbr a||{\bm t}^{(1)}||b\rbr\,,
\end{equation}
where $a$ and $b$ numerate the one-electron orbitals and $d_{rs}^{(1)}$ are the
angular recoupling coefficients. Packages for the numerical evaluation of the
coefficients $d_{rs}^{(1)}$ are available in the literature
\cite{pyper:78:cpc,gaigalas:02:cpc}. The reduced matrix element of the
one-electron operator ${\bm t}^{(1)}$ is given by
\begin{align}\label{2eq5}
    \lbr a||{\bm t}^{(1)}||b\rbr = &\, -(\kappa_a+\kappa_b)\, \lbr
    -\kappa_a||{\bm{C}}^{(1)}||\kappa_b\rbr\,
 \nonumber \\ & \times
    \int_0^{\infty}dr\,
      r^{-2}\,(G_aF_b+F_aG_b)\,.
\end{align}

The energy shift due to the hfs splitting can easily be calculated for the
hydrogen-like ion. In this case, the sum (\ref{2eq4}) consists of a single term
and the radial integral in Eq.~(\ref{2eq5}) is calculated analytically (in the
point-nucleus limit). In the present work, we will need the nonrelativistic limit
of Eq.~(\ref{2eq5}) for the hydrogen-like ion, which reads
\begin{equation}\label{2eq6}
    \lbr njl||{\bm t}^{(1)}||njl\rbr_{NR} = \frac{2(\Za)^3\,m^2}{n^3}\,\frac{1}{2l+1}\,
        \sqrt{\frac{2j+1}{j(j+1)}}\,.
\end{equation}
Using this result, it is convenient to introduce the following parametrization of
the magnetic hyperfine constant $A_J$:
\begin{equation}\label{2eq7}
    A_J = \frac{\alpha(\Za)^3}{n^3}\,
        \frac{m^2}{m_p}\,\frac{\mu}{\mu_N}\,
      \frac{1}{I J(J+1)(2L+1)}\, G_{M1}(Z)\,,
\end{equation}
where $n$ is the principal quantum number of the valence electron, $m_p$ is the
proton mass, and $\mu_N = |e|/(2m_p)$ is the nuclear magneton. The function
$G_{M1}(Z)$ is dimensionless; its numerical value is unity for a hydrogen-like
nonrelativistic atom in the point-nucleus and non-recoil limit. $G_{M1}$ is a
slowly varying function of the nuclear charge number $Z$ and the quantum numbers
$J$ and $L$, which is convenient for the representation of numerical results. This
definition of the function $G_{M1}$ differs slightly from the one used in our
previous work \cite{yerokhin:08:pra} by the fact that it does not include the
nonrelativistic mass scaling factor $(1+m/M)^{-3}$. We presently choose to treat
this part of the recoil effect (also referred to as the normal mass shift) on an
equal footing with the other recoil corrections. A parametrization similar to that
in Eq.~(\ref{2eq7}) was previously used in
Refs.~\cite{shabaev:94:hfs,shabaev:97:pra}.

Numerical results of nonrelativistic calculations are often presented in terms of
the orbital ($a_l$), the spin-dipole ($a_{sd}$), and the Fermi contact ($a_c$)
hyperfine parameters, induced by the three terms in Eq.~(\ref{2eq2b}) and defined
as \cite{hibbert:75}
\begin{align}\label{2eq11}
    a_l &\ = \lbr LSM_LM_S|\sum_{i=1}^N \frac{l_0^{(1)}(i)}{r_i^{3}}
    |LSM_LM_S\rbr\,, \\
    a_{sd} &\ = \lbr LSM_LM_S|\sum_{i=1}^N \frac{2 C_0^{(2)}(i)\,s_0^{(1)}(i)}{r_i^{3}}
    |LSM_LM_S\rbr\,, \\
    a_{c} &\ = \lbr LSM_LM_S|\sum_{i=1}^N \frac{2 s_0^{(1)}(i)\, \delta(r_i)}{r_i^{2}}
    |LSM_LM_S\rbr\,, \\
\end{align}
with $M_L=L$ and $M_S=S$. The connection of the (nonrelativistic limit of the)
function $G_{M1}$ with the hyperfine parameters expressed in atomic units is given
by
\begin{equation}\label{2eq8}
    {G_{M1}}(Z) = \frac{n^3
    J(J+1)(2L+1)}{2 Z^3}\,(c_l\,a_l+c_{sd}\,a_{sd}+c_c\,a_c)\,.
\end{equation}
The coefficients $c_i$ are \cite{hibbert:75}
\begin{eqnarray} \label{2eq9}
  c_l &=& \frac{\lbr \bm{L}\cdot\bm{J}\rbr}{LJ(J+1)}\,, \\
  c_{sd} &=& \frac{3\lbr \bm{S}\cdot\bm{L}\rbr\, \lbr \bm{L}\cdot\bm{J}\rbr
       -L(L+1)\lbr \bm{S}\cdot\bm{J}\rbr}
           {SL(2L-1)J(J+1)} \,,\\
  c_c &=& \frac{\lbr \bm{S}\cdot\bm{J}\rbr}{3SJ(J+1)}\,,
\end{eqnarray}
where
\begin{eqnarray} \label{2eq10}
\lbr \bm{L}\cdot\bm{J}\rbr &=& [J(J+1)+L(L+1)-S(S+1)]/2\,, \\
\lbr \bm{S}\cdot\bm{J}\rbr &=& [J(J+1)-L(L+1)+S(S+1)]/2\,, \\
\lbr \bm{S}\cdot\bm{L}\rbr &=& [J(J+1)-L(L+1)-S(S+1)]/2\,.
\end{eqnarray}

\subsection{Electric quadrupole hyperfine splitting}

The scalar part of the interaction between an electron and the nucleus is given by
\begin{equation}\label{3eq1}
    V(\bfr,\bfr_{p_1},\ldots,\bfr_{p_Z}) = -\alpha \sum_{j = 1}^Z
    \frac1{|\bfr-\bfr_{p_j}|}\,,
\end{equation}
where $\bm{r}$ and $\bfr_{p_j}$ are the coordinates of the electron and the $j$th
proton, respectively. Averaging this interaction over the internal nuclear
coordinates and using the standard multipole expansion of
$|\bfr-\bfr_{p_j}|^{-1}$, one obtains (see Ref.~\cite{kozhedub:08} for the
details)
\begin{align}\label{3eq2}
 V_{\rm av}(\bfr,\Theta,\Phi) &\, \equiv \lbr V \rbr_{\rm intern}
  = -\alpha \sum_{l = 0}^{\infty}
    \int_0^{\infty}dr\pr\,{r\pr}^2\, \rho_l(r\pr)\,
 \nonumber \\ & \times
       \left[ \frac{r^l}{{r\pr}^{l+1}}\theta(r\pr-r)
         +\frac{{r\pr}^l}{{r}^{l+1}}\theta(r-r\pr)\right]\,
 \nonumber \\ & \times
            {\bm{C}}^{(l)}(\hr)\cdot {\bm{C}}^{(l)}(\Theta,\Phi)\,,
\end{align}
where $\lbr\cdots\rbr$ denotes the averaging, $\Theta$ and $\Phi$ are the angles
that fix the orientation of the intrinsic nuclear system with respect to the
laboratory frame, and the nuclear charge density component $\rho_l$ is defined as
\begin{equation}\label{3eq3}
    \rho_l(r) = \int d\hr\, \rho(\bfr)\,C^{(l)}_0(\hr)\,.
\end{equation}

The first term in Eq.~(\ref{3eq2}) ($l=0$) yields the standard Coulomb interaction
between the electron and the nucleus with an extended charge distribution. The
term with $l=1$ vanishes after averaging with the electron wave function of a
definite parity. The term with $l=2$ gives rise to a splitting of the energy level
(of an electronic state with $J>1/2$), known as the electric quadrupole one. The
corresponding interaction is conveniently written in the form
\begin{equation}\label{3eq4}
    H_{E2} = \alpha\,\bm{T}^{(2)}\cdot\bm{Q}^{(2)}_{\rm av}\,.
\end{equation}
Here, $\bm{Q}^{(2)}_{\rm av}$ is the operator of the nuclear quadrupole moment
averaged over the internal (radial) nuclear coordinates,
\begin{equation}\label{3eq5}
  \bm{Q}^{(2)}_{\rm av} =
\lbr \bm{Q}^{(2)} \rbr_{\rm intern} = N\,{\bm{C}}^{(2)}(\Theta,\Phi)\,.
\end{equation}
The normalization constant $N$ is
\begin{equation}\label{3eq6}
    N = \int_0^R dr\, r^4\,\rho_2(r)\,,
\end{equation}
where $R$ is the nuclear radius. The operator $\bm{T}^{(2)}$ acts on the
electronic variables. It is given by
\begin{equation}\label{3eq7}
  \bm{T}^{(2)} = \sum_i \bm{t}^{(2)}(i) = -\sum_i f(r_i)\,{\bm{C}}^{(2)}(\hr_i)\,,
\end{equation}
where the radial distribution function $f(r)$ is
\begin{equation}\label{3eq7a}
    f(r) = \left\{
       \begin{array}{l}
        \displaystyle
           \frac1{r^3}\,,\ \ r>R\,, \\
        \displaystyle
            \frac1N
              \int_0^R dr\pr {r\pr}^2\,\rho_2(r\pr)\, \frac{r_<^2}{r_>^3}
               \,,
               \ \ r\le R\,.
       \end{array}
       \right.
\end{equation}
The distribution function $f(r)$ can easily be calculated analytically for several
simple models of the nuclear-charge distribution. So, if $\rho_2$ does not depend
on $r$ within the nucleus, $\rho_2(r) \propto \theta(R-r)$,
\begin{equation}\label{3eq7b}
    f(r) = \frac{r^2}{R^5}\left( 1+ 5\ln \frac{R}{r} \right)\,,\ \  r\le R\,.
\end{equation}
If $\rho_2(r)  \propto\delta(R-r)$, then
\begin{equation}\label{3eq7c}
    f(r) = \frac{r^2}{R^5}\,,\ \  r\le R\,.
\end{equation}
In the point-quadrupole limit, the function $f(r)$ takes the standard form, $f(r)
= r^{-3}$.

The finite nuclear size effect is very small for the electric quadrupole splitting
and its inclusion in calculations is not necessary at present. However, we
observed that the usage of the extended charge distribution considerably improves
the stability and the convergence of numerical calculations. The reason for this
is that the extended distribution removes the $r^{-3}$ singularity of the
point-quadrupole interaction.

Using the standard technique of the angular-momentum algebra (see, e.g.,
Ref.~\cite{hibbert:75}), the energy shift due to the electric quadrupole
interaction can be expressed in terms of the reduced matrix elements of the
electronic operator $\bm{T}^{(2)}$. The correction to the energy is usually
expressed in terms of the hyperfine structure constant $B_J$, which does not
depend on the total angular momentum of the system $F$,
\begin{equation}\label{3eq10}
    \Delta E_{E2} = \frac{\frac34\, C(C+1)-I(I+1)J(J+1)}{2I(2I-1)J(J+1)}\, B_J\,,
\end{equation}
where $C = F(F+1)-I(I+1)-J(J+1)$ and
\begin{equation}\label{3eq11}
    B_J = 2Q\,\sqrt{\frac{2J(2J-1)}{(2J+1)(2J+2)(2J+3)}}\,
       \lbr J||{\bm T^{(2)}}||J\rbr\,.
\end{equation}
Here, $Q$ is the nuclear quadrupole moment, defined as
\begin{equation}\label{3eq12}
    Q =  \lbr IM | \sum_{j=1}^Z(3z_{p_j}^2-r_{p_j}^2)|IM\rbr_{M=I}
     =  2\,\lbr II | {\bm Q}_{0}^{(2)}|II\rbr\,.
\end{equation}

The reduced matrix element of the operator ${\bm T}^{(2)}$ should be evaluated
with the many-electron wave functions (\ref{eq4}), obtained by the solution of the
secular equation (\ref{eq6}). Matrix elements of the operator ${\bm T}^{(2)}$
between individual CSFs can be expressed as linear combinations of matrix elements
of the one-electron operator ${\bm t}^{(2)}$ between the single-particle orbitals,
\begin{equation}\label{3eq13}
  \lbr \gamma_r PJ||{\bm T}^{(2)}||\gamma_s PJ\rbr = \sum_{a \le b}
  \,d_{rs}^{(2)}(ab)\, \lbr a||{\bm t}^{(2)}||b\rbr\,,
\end{equation}
where $a$ and $b$ numerate the one-electron orbitals and $d_{rs}^{(2)}$ are the
angular recoupling coefficients \cite{pyper:78:cpc}. The reduced matrix element of
the one-electron operator ${\bm t}^{(2)}$ is given by
\begin{equation}\label{3eq14}
    \lbr a||{\bm t}^{(2)}||b\rbr = - \lbr
    \kappa_a||{\bm{C}}^{(2)}||\kappa_b\rbr\, \int_0^{\infty}dr\,
      f(r)\,(G_aG_b+F_aF_b)\,.
\end{equation}

Similarly to the magnetic dipole hyperfine constant $A_J$, the electric quadrupole
hyperfine constant $B_J$ can be conveniently parameterized by introducing the
dimensionless function $G_{E2}$, which turns into unity for a nonrelativistic
hydrogen-like ion in the point-nucleus and non-recoil limit,
\begin{equation}\label{3eq15}
    B_J = Q\, \frac{\alpha (\Za)^3m^3}{n^3} \,\frac{2J-1}{J+1}\,
           \frac{1}{\kappa(\kappa+1)(2L+1)}\, G_{E2}(Z)\,,
\end{equation}
where $\kappa = (-1)^{J+L+1/2}(J+1/2)$.

Results of nonrelativistic calculations are often expressed in terms of the
quadrupole parameter $b_q$, defined as
\begin{equation}\label{3eq16}
    b_q = \lbr LSM_LM_S|\sum_{i=1}^N \frac{2 C_0^{(2)}(\hr_i)}{r_i^{3}}
    |LSM_LM_S\rbr_{M_L=L,M_S=S}\,.
\end{equation}
The connection between the (nonrelativistic limit of the) function $G_{E2}$ and
the parameter $b_q$ expressed in atomic units is given by
\begin{equation}\label{3eq17}
    G_{E2}(Z) = \frac{n^3(J+1)\kappa(\kappa+1)(2L+1)}{Z^3(2J-1)}\,(-c_q\,b_q)\,,
\end{equation}
where the coefficient $c_q$ is \cite{hibbert:75}
\begin{equation}\label{3eq18}
    c_q = \frac{6\lbr \bm{L}\cdot\bm{J}\rbr^2
    -3\lbr\bm{L}\cdot\bm{J}\rbr-2L(L+1)J(J+1)}{L(2L-1)(2J+3)(J+1)}\,.
\end{equation}

%
%
\subsection{Details of the CI calculation}

To perform a CI calculation, we devised a code, incorporating and adapting a
number of existing packages
\cite{grant:73:cpc,pyper:78:cpc,grant:80:cpc,stathopoulos:94:cpc,gaigalas:01:cpc}
for setting up the CSFs, calculating angular-momentum coefficients, and
diagonalizing the Hamiltonian matrix. The largest number of CSFs simultaneously
handled was about a half of a million. A careful optimization of the code was
necessary to keep the time and memory consumption of the calculation within
reasonable limits. Care was taken to prevent re-calculating the angular-momentum
coefficients for the pairs of CSFs that differ by the principal quantum number of
a single electron only. An optimized ordering of CSFs allows one to drastically
reduce the number of angular-momentum coefficients to be evaluated. A similar
optimization was introduced in the calculation of the Coulomb and Breit radial
integrals. The radial integrals with the same pair of electron states in the
innermost radial integration were grouped together and evaluated simultaneously.

The dominant part of the hfs splitting is delivered by the Dirac-Coulomb
Hamiltonian. This is the most demanding part of the calculation, since a high
relative precision is required. One of the factors defining the accuracy of the
calculation is the quality and the size of the space of one-electron orbitals from
which the CSFs are constructed. We take this space to be a part of the finite
basis set of eigenvectors of the Dirac equation, obtained by the
dual-kinetic-balance method \cite{shabaev:04:DKB} and constructed with $B$-splines
\cite{deboor:78}.

For a given number of B-splines $n_a$, all eigenstates were taken with the energy
$0< \vare \le mc^2(1+Z\alpha\, E_{\rm max})$ and the orbital quantum number $l \le
l_{\rm max}$, where the value of $E_{\rm max}$ was varied between $0.5$ and $6$
and $l_{\rm max}$, between $1$ and $7$. Three main sets of one-electron orbitals
were employed in the present work: (A) $20s\,$$20p\,$$19d\,$$19f\,$$18g\,$$18h$
with $n_a=44$ and $E_{\rm max}=3.0$, (B)
$14s\,$$14p\,$$14d\,$$13f\,$$13g\,$$13h\,$$12i\,$$12k$ with $n_a=34$ and $E_{\rm
max}=0.5$,  and (C) $25s\,$$25p\,$$24d\,$ with $n_a=54$ and $E_{\rm max}=6.0$.
Here, the notation, e.g., $20p$ means $20p_{1/2}\,$$20p_{3/2}$. Calculational
results were first obtained with the set (A) and then corrected for contributions
of the higher partial waves with the set (B) and for a more complete
representation of the Dirac spectrum with the set (C). The computation became
rather intensive for the $P$ states, so the basis set (A) was reduced to include
the states with $l\leq 3$ only in this case. Usage of several sets of one-electron
orbitals allowed us to efficiently control the completeness of the representation
of the Dirac spectrum in our calculations.

The analysis of the convergence of the partial-wave expansion was performed by
identifying increments of the results induced by the increasing cutoff parameter
$l_{\rm max}$. The omitted tail of the expansion was estimated by a polynomial
least-square fitting of the increments in $1/l$. In most cases, the error due to
the termination of the expansion was found to yield the largest uncertainty to the
Dirac-Coulomb hfs value.

The set of the CSFs employed in the calculation was obtained by taking all single,
double, and triple excitations of the reference configuration with at least one
electron orbital with $l\leq 1$ present. The contribution of the remaining triple
excitations was found to be negligible for the $S$ states. For the $P$ states, it
was estimated by repeating the calculation with a smaller basis but with the above
restriction replaced by $l\leq 2$.

Inclusion of the Breit interaction into the Dirac-Coulomb Hamiltonian yields only
a small correction in the case of Li and Be$^+$. Because of this, it is sufficient
to use much smaller basis sets for its evaluation, which simplifies the
computation greatly. The Breit-interaction correction was obtained as the
difference of the CI results with and without the Breit interaction included into
the Hamiltonian, evaluated with the same set of CSFs.

Results of our CI calculations of the magnetic dipole and the electric quadrupole
hfs splitting are presented in Tables~\ref{tab:dipole} and \ref{tab:quadrupole},
respectively. The CI values obtained with the Dirac-Coulomb Hamiltonian are listed
under the entry ``Coulomb''; the entry ``Breit'' contains the correction due to
the inclusion of the Breit interaction. The comparison presented in the tables
demonstrates significant deviations of our CI values from the MCDF results by
Biero\'n {\em et al.} \cite{bieron:96,bieron:99} and from the CC results of
Johnson {\em et al.} \cite{johnson:08}. In the case of Be$^+$,
Ref.~\cite{bieron:99} reports estimations of the calculational errors, considered
by the authors to be the conservative ones, but our CI results are well out of
these error bars for all the states studied.

The deviation from the MCDF calculations is the strongest for the quadrupole
splitting in Be$^+$.  In this case, our CI value differs from the MCDF one already
in the second digit, while the claimed accuracy of the MCDF result is about
$10^{-5}$. A similar deviation is observed also for the quadrupole splitting in
Li. At the same time, agreement of our calculations with the nonrelativistic
studies \cite{godefroid:01,guan:98} is much better, on the level of $10^{-3}$.
This observation leads us to a conclusion that the MCDF results for the quadrupole
splitting are, most probably, in error. A possible explanation for this is that
the highly singular point-quadrupole interaction might lead to considerable
numerical errors when evaluated on approximate relativistic wave functions. In our
calculations, we detected such problems; they were solved by using the extended
charge distribution for the quadrupole interaction.

In order to make possible a detailed comparison with high-precision
nonrelativistic results available in the literature, we have to identify the
nonrelativistic part of our CI values. This was achieved by repeating the full set
of the CI calculations for different values of the fine-structure constant
$\alpha$ (namely, three values with ratios $\alpha^{\prime}/\alpha = 0.9$, $1$,
and $1.1$ were used). For each value of $\alpha$, the finite nuclear-charge
distribution correction was evaluated (as described in the next section) and
subtracted from the CI values. The point-nucleus results thus obtained were fitted
to a polynomial in $\alpha$, assuming the absence of the linear term. In this way,
the CI results with the physical value of $\alpha$ were separated into three
parts: the nonrelativistic point-nucleus contribution, the relativistic
correction, and the finite nuclear-charge correction.

\setlength{\LTcapwidth}{\textwidth}
\begingroup
\begin{ruledtabular}
\begin{longtable*}{ll....l}
\caption{Individual contributions to the magnetic dipole hfs splitting in
$^{6,7}$Li and $^9$Be$^+$, in terms of the function $G_{M1}$ if not specified
otherwise. For $^{6}$Li, only the contributions different from those for $^{7}$Li
are listed. The values of the nuclear magnetic moments are taken from
Ref.~\cite{stone:05}. The entries are labelled as follows: ``NR(point)'' denotes
the point-nucleus nonrelativistic result; ``Relativistic'' is the total
relativistic correction; ``Coulomb'' is the relativistic hfs value obtained with
the Dirac-Coulomb Hamiltonian; ``Breit'' is the Breit-interaction correction;
``BW'' is the nuclear magnetization distribution correction; ``NMS'' is the normal
mass shift; ``SMS'' is the specific mass shift; ``SO'' is the sum of the normal
and specific spin-orbital recoil corrections induced by Eqs.~(\ref{4eq6b}) and
(\ref{4eq6bb}); ``Negative-energy'' is the contribution of the negative-energy
part of the Dirac spectrum.
 \label{tab:dipole}}
\\
\hline \hline\\[-5pt]
    &  &\multicolumn{1}{c}{$2^2S$}  &\multicolumn{1}{c}{$3^2S$}  &\multicolumn{1}{c}{$2^2P_{1/2}$}
                                                                      &\multicolumn{1}{c}{$2^2P_{3/2}$} & \mbox{Ref.}
\\
\colrule
\endfirsthead
\\[-5pt]
\multicolumn{7}{c}{$^7$Li}\\[5pt]
NR(point)        &  &   0.215x251        &    0.168x340        &     0.073x905       &   -0.024x348     & \\
                 &  &   0.215x254\,(4)   &    0.168x351\,(13)  &                     &                  & Hylleraas \cite{yan:96} \\
                 &  &   0.215x19         &    0.168x28         &     0.073x89        &   -0.024x51      & MCHF \cite{godefroid:01} \\
Relativistic     &  &   0.000x205        &    0.000x159        &     0.000x018       &   -0.000x036     & \\
Coulomb          &  &   0.215x385\,(5)   &    0.168x440\,(9)   &     0.073x923\,(1)  &   -0.024x364\,(10)& \\
                 &  &   0.215x27         &                     &     0.073x96        &   -0.024x76      & MCDF \cite{bieron:96} \\
                 &  &   0.215x65         &    0.168x61         &     0.073x89        &   -0.024x25      & CCSD \cite{johnson:08} \\
Breit            &  &   0.000x016        &    0.000x016        &    -0.000x003       &    0.000x000     & \\
QED              &  &   0.000x182\,(4)   &    0.000x143\,(3)   &     0.000x048\,(1)  &   -0.000x085\,(2)& \\
BW               &  &  -0.000x024\,(5)   &   -0.000x019\,(4)   &    -0.000x002       &    0.000x009\,(2)& \\
Recoil        & NMS &  -0.000x050        &   -0.000x039        &    -0.000x017       &    0.000x006     & \\
              & SMS &   0.000x002        &    0.000x002        &     0.000x027       &   -0.000x055     & \\
              & SO  &   0.000x000        &    0.000x000        &    -0.000x001\,(1)  &   -0.000x002\,(1)& \\
Negative-energy  &  &   0.000x002\,(1)   &    0.000x002\,(1)   &    -0.000x003\,(1)  &   -0.000x003\,(2)& \\[5pt]
Total            &  &   0.215x512\,(8)   &    0.168x544\,(11)  &     0.073x972\,(2)  &   -0.024x493\,(10)& \\
Total$^a$ [MHz]  &  &   401.7x55\,(15)   &    93.09x5\,(6)     &     45.96x6\,(1)    &   -3.044x\,(1)   & \\
Experiment [MHz] &  &   401.7x520433(5)^b&    93.10x6\,(11)^c  &     45.91x4\,(25)^d &   -3.055x\,(14)^d& \\
                 &  &                    &                     &     46.01x0\,(25)^e &                  & \\
                 &  &                    &                     &     46.02x4\,(3)^f  &                  & \\
[5pt]%
\multicolumn{7}{c}{$^6$Li}\\[5pt]
Coulomb          &  &   0.215x382\,(5)   &    0.168x438\,(9)   &     0.073x922\,(1)  &   -0.024x363\,(10)& \\
BW               &  &  -0.000x022\,(13)  &   -0.000x017\,(10)  &    -0.000x002\,(1)  &    0.000x008\,(5) & \\
Recoil       &  NMS &  -0.000x059        &   -0.000x046        &    -0.000x020       &    0.000x007      & \\
             &  SMS &   0.000x002        &    0.000x002        &     0.000x031       &   -0.000x064      & \\[5pt]
Total            &  &   0.215x504\,(14)  &    0.168x538\,(14)  &     0.073x974\,(2)  &   -0.024x502\,(11)& \\
Total$^g$ [MHz]  &  &   152.1x22\,(10)   &    35.25x0\,(3)     &     17.40x58\,(5)   &   -1.153x0\,(5)   & \\
Experiment [MHz] &  &   152.1x36839\,(2)^b&   35.26x3\,(15)^c  &     17.37x5\,(18)^h &   -1.155x\,(8)^h  & \\
                 &  &                    &                     &     17.38x6\,(31)^e &                   & \\
                 &  &                    &                     &     17.39x4\,(4)^f  &                   & \\
[5pt]%
\multicolumn{7}{c}{$^9$Be$^+$}\\[5pt]
NR(point)        &  &   0.390x544        &    0.335x066        &     0.221x132       &    0.009x89    & \\
                 &  &   0.390x549\,(9)   &                     &                     &                & Hylleraas \cite{yan:96} \\
                 &  &   0.390x50         &    0.335x04         &     0.221x13        &    0.009x67    & MCHF \cite{godefroid:01} \\
Relativistic     &  &   0.000x664        &    0.000x563        &     0.000x162       &   -0.000x15    & \\
Coulomb          &  &   0.391x030\,(6)   &    0.335x468\,(9)   &     0.221x302\,(2)  &    0.009x800\,(25)    & \\
                 &  &   0.390x94\,(4)    &                     &     0.221x40\,(1)   &    0.009x1\,(4) & MCDF \cite{bieron:99} \\
Breit            &  &   0.000x039        &    0.000x042        &    -0.000x021       &   -0.000x001        & \\
QED              &  &   0.000x289\,(12)  &    0.000x248\,(10)  &     0.000x137\,(5)  &   -0.000x181\,(7)    & \\
BW               &  &  -0.000x062\,(6)   &   -0.000x053\,(5)   &    -0.000x005\,(1)  &    0.000x027\,(3)    & \\
Recoil        & NMS &  -0.000x071        &   -0.000x061        &    -0.000x040       &   -0.000x002        & \\
              & SMS &   0.000x002        &    0.000x002        &     0.000x057       &   -0.000x080        & \\
              & SO  &   0.000x000        &    0.000x000        &    -0.000x007\,(2)  &   -0.000x018\,(5)   & \\
Negative-energy  &  &   0.000x005\,(3)   &    0.000x005\,(2)   &    -0.000x009\,(4)  &   -0.000x011\,(6)    & \\[5pt]
Total            &  &   0.391x233\,(15)  &    0.335x651\,(15)  &     0.221x413\,(7)  &    0.009x533\,(27)   & \\
Total$^i$ [MHz]  &  &  -625.0x8\,(2)     &   -158.8x97\,(7)    &    -117.9x19\,(4)   &   -1.015x\,(3)       & \\
Total$^j$ [MHz]  &  &  -625.1x1\,(3)     &   -158.9x05\,(7)    &    -117.9x25\,(4)   &   -1.016x\,(3)       & \\
Experiment [MHz] &  &  -625.0x0883705\,(1)^k&                  &    -118.6x\,(36)^l  &                     & \\
\hline\hline
$^a $ $\mu(^7{\rm Li}) = 3.2564268\,(17)$, \\
$^b $ Beckmann {\em et al.}, 1974 \cite{beckmann:74}, \\
$^c $ Bushaw {\em et al.}, 2003 \cite{bushaw:03}, \\
$^d $ Orth {\em et al.}, 1975 \cite{orth:75}, \\
$^e $ Walls {\em et al.}, 2003 \cite{walls:03}, \\
$^f $ Das and Natarajan, 2008 \cite{das:08}, \\
$^g $ $\mu(^6{\rm Li}) = 0.8220473\,(6)$, \\
$^h $ Orth {\em et al.}, 1974 \cite{orth:74},\\
$^i $ $\mu(^9{\rm Be}) = -1.177432\,(3)$, \\
$^j $ $\mu(^9{\rm Be}) = -1.177492\,(17)$,  \\
$^k $ Wineland {\em et al.}, 1983 \cite{wineland:83:prl}, \\
$^l $ Bollinger {\em et al.}, 1985 \cite{bollinger:85}.\\
\end{longtable*}
\end{ruledtabular}
\endgroup


\begin{table*}
\setlength{\LTcapwidth}{\textwidth}
\caption{Individual contributions to the electric quadrupole hfs splitting of the
$2^2P_{3/2}$ state, in terms of the function $G_{E2}$ if not specified otherwise.
The notations are the same as in Table~\ref{tab:dipole}. The values of the nuclear
quadrupole moments are taken from Ref.~\cite{stone:05}.
 \label{tab:quadrupole}}
\begin{ruledtabular}
\begin{tabular}{l...l}
    &\multicolumn{1}{c}{$^7$Li}&\multicolumn{1}{c}{$^6$Li}&\multicolumn{1}{c}{$^9$Be$^+$}  & Ref.\\
\hline\\
NR               &    0.050x260       &    0.050x260         &     0.172x140     & \\
                 &    0.049x8         &                      &     0.171x7       & MCHF \cite{godefroid:01}\\
                 &    0.049x8         &                      &     0.172x7       & FCPC \cite{guan:98}\\
Relativistic     &   -0.000x004       &   -0.000x004         &    -0.000x013     & \\
Coulomb          &    0.050x260\,(3)  &    0.050x260\,(3)    &     0.172x150\,(7) & \\
                 &    0.051x085       &                      &     0.183x56\,(3)  & MCDF \cite{bieron:96,bieron:99}\\
Breit            &   -0.000x004       &   -0.000x004         &    -0.000x024     & \\
QED              &    0.000x000\,(2)  &    0.000x000\,(2)    &     0.000x000\,(11)& \\
NMS              &   -0.000x012       &   -0.000x014         &    -0.000x031     & \\
SMS              &    0.000x012       &    0.000x014         &     0.000x030     & \\
Total            &    0.050x256\,(4)  &    0.050x256\,(4)    &     0.172x125\,(13)& \\
Total [MHz]      &   -0.216x\,(4)^a   &    -0.004x4\,(1)^b   &      2.281x\,(16)^c  & \\
Experiment [MHz] &   -0.221x\,(29)    &    -0.010x(14)        &            &  \cite{orth:74,orth:75}\\
\end{tabular}
$^a$  $Q(^7{\rm Li}) = -40.55\,(80)$~mbarn,\\
$^b$  $Q(^6{\rm Li}) = -0.82\,(2)$~mbarn,\\
$^c$  $Q(^9{\rm Be}) =  52.88\,(38)$~mbarn,
\end{ruledtabular}
\end{table*}

\begin{table*}
\setlength{\LTcapwidth}{\textwidth}
\caption{Non-relativistic hfs parameters, in a.u.
 \label{tab:NR}}
\begin{ruledtabular}
\begin{tabular}{c......l}
Ion &\multicolumn{1}{c}{$2^2S$}  &\multicolumn{1}{c}{$3^2S$}  &\multicolumn{4}{c}{$2^2P$} & Ref. \\
    &\multicolumn{1}{c}{$a_c$}   &\multicolumn{1}{c}{$a_c$}  &\multicolumn{1}{c}{$a_c$}
             &\multicolumn{1}{c}{$a_{sd}$}&\multicolumn{1}{c}{$a_l$}&\multicolumn{1}{c}{$b_q$} & \\
\hline\\[-9pt]
Li   &     2.90x589    &  0.67x336    &   -0.21x467    &  -0.01x3477  &   0.06x3125  &  -0.02x2617 & This work \\
     &     2.90x592(5) &  0.67x341(5) &   -0.21x478(5) &              &              &             & Hylleraas \cite{yan:96} \\
     &     2.90x51     &  0.67x31     &   -0.21x51     &  -0.01x346   &   0.06x311   &  -0.02x239  & MCHF \cite{godefroid:01} \\
     &     2.90x3      &  0.67x45     &   -0.21x36     &  -0.01x341   &   0.06x309   &  -0.02x242  & FCPC \cite{guan:98} \\
[9pt]%
Be$^+$&   12.4x974     & 3.17x69      &  -1.08x42      & -0.10x269   &    0.48x520   &  -0.18x362  & This work \\
      &   12.4x976(3)  &              &                &             &               &             & Hylleraas \cite{yan:96} \\
      &   12.4x96      & 3.17x67      &  -1.08x56      & -0.10x265   &    0.48x516   &  -0.18x310  & MCHF \cite{godefroid:01} \\
      &   12.4x93      & 3.18x1       &  -1.07x0       & -0.10x20    &    0.48x51    &  -0.18x42   & FCPC \cite{guan:98}  \\
\end{tabular}
\end{ruledtabular}
\end{table*}

For the $P$ states and the magnetic dipole hfs, the nonrelativistic limit of the
CI results needs to be separated into three parts, corresponding to the three
terms of the nonrelativistic decomposition of the hfs operator (\ref{2eq2b}). To
this end, we carried out identical calculations both for the relativistic magnetic
dipole hfs operator (\ref{2eq2}) and for the spin-dipole and the orbital parts of
its nonrelativistic decomposition. Applying the fitting procedure described above,
we identify the nonrelativistic limit of the CI values as well as the spin-dipole
and orbital hfs parameters $a_l$ and $a_{sd}$. The remaining contact parameter
$a_c$ is then unambiguously deduced. (We prefer not to perform a direct
calculation for the contact term since the corresponding operator contains a
$\delta$-function and needs a regularization when evaluated on relativistic wave
functions.)

The nonrelativistic hfs parameters obtained in this way are listed in
Table~\ref{tab:NR}. The nonrelativistic results for the $2^2P$ state were obtained
from the relativistic calculations for the $2^2P_{1/2}$ state. Since in the
present work the hfs parameters are needed for the purpose of comparison only, we
do not assign the uncertainty to them (which is difficult to do reliably since
they are obtained by a fit). The comparison with the previous nonrelativistic
calculations \cite{yan:96,godefroid:01,guan:98} presented in the table exhibits a
remarkably good agreement of our values with the high-precision results obtained
in a Hylleraas-type calculation by Yan {\em et al.} \cite{yan:96}.

In Tables~\ref{tab:dipole} and \ref{tab:quadrupole}, the entry "NR(point)" labels
the nonrelativistic, point-nucleus limits of the functions $G_{M1}$ and $G_{E2}$
obtained by the fitting procedure described above. Because of the fitting, the
uncertainties are not ascribed to them; we expect that they are somewhat less
accurate than the corresponding relativistic values. The comparison is drawn with
the most accurate previous nonrelativistic calculations. A much better agreement
is observed with the previous nonrelativistic results than with the relativistic
ones.

\section{Corrections to the hyperfine splitting}
\label{sec:corrections}

While the evaluation of the relativistic hfs value is the most computationally
intensive part of the calculation, a high-precision theoretical determination of
the hfs splitting requires inclusion of a number of important corrections. In this
section, we present a detailed description of each of them in turn.

\subsection{QED effects}

For the magnetic dipole hfs splitting, the leading (in $\Za$) QED contribution
originates from the anomalous magnetic moment of the electron $g_e$. The effect is
accounted for by multiplying the spin-dependent terms in Eq.~(\ref{2eq2a}) by
$g_e/2 \approx \alpha/(2\pi)$, see, e.g., Ref.~\cite{hibbert:75}. So, the leading
QED correction to the function $G_{M1}$ is given by
\begin{equation}\label{4eq1}
    \delta G_{M1}^{QED,0}(Z) = \frac{\alpha}{2\pi}\, \Bigl[ G_{M1,sd}(Z)+
             G_{M1,c}(Z)\Bigr]\,,
\end{equation}
where $G_{M1,sd}$ and $G_{M1,c}$ are the contributions to the function $G_{M1}$
induced by the spin-dipole and the contact term in Eq.~(\ref{2eq2a}),
respectively.

The higher-order terms of the $\Za$ expansion (the {\em binding} QED corrections)
induce important contributions and should be taken into account alongside with the
leading effect. The binding corrections to the contact term can be written in a
form analogous to that for the hydrogen hfs
\cite{sapirstein:90:kin,yan:96,pachucki:02:pra},
\begin{align}\label{4eq2}
    \delta G_{M1}^{QED,bind}&\,(Z) = \frac{\alpha}{\pi}\, G_{M1,c}(Z)
    \left\{ \Za \,\pi \left(\ln 2-\frac52 \right)
   \right.
 \nonumber \\ &
   \left.
      +(\Za)^2 \Bigl[ -\frac83\,\ln^2(\Za)
      +a_{21}\,\ln(\Za)+a_{20} \Bigr] \right\}\,.
\end{align}
The coefficients $a_{21}$ and $a_{20}$ are different from the hydrogenic case and
not known at present. One can, however, use their hydrogenic values as crude
estimates. In our calculations, we will use the results for the hydrogenic $2s$
state, $a_{21} = -1.1675$, $a_{20} = 11.3522$
\cite{sapirstein:90:kin,karshenboim:02:epjd}, and assume a 100\% uncertainty for
them. This treatment of the QED effects coincides with those of
Refs.~\cite{yan:96,king:07} but is different from other previous investigations,
where the binding effects were continually neglected. Such neglect can hardly be
justified since the higher-order terms change the total QED contribution by 40\%
for lithium and by 60\% for beryllium.

The binding corrections to the spin-dipole and orbital parts of hfs are relevant
for the states with $l>0$ only. They enter in the relative order $\alpha(\Za)^2$
and are presently unknown. Numerical calculations for the hydrogenic case
\cite{sapirstein:06:hfs} show that their nominal order can be enhanced by the
second power of logarithm. We thus estimate the uncertainty due to their neglect
by
$$
|G_{M1,sd}+G_{M1,l}|\, \frac{\alpha}{\pi}\, (\Za)^2\, \ln^2(\Za)\,.
$$

For the electric quadrupole splitting, the QED correction has not been calculated
so far. Its relative nominal order is $\alpha(\Za)^2$. According to our analysis,
this correction diverges in the point-quadrupole limit $R\to 0$, which means that
the nominal order is enhanced by $\ln R \approx 6$. We, therefore, estimate the
error in $G_{E2}$ due to the neglect of the QED effects by multiplying it by the
factor of $$10\,\alpha (\Za)^2\,.$$

\subsection{Nuclear recoil}

Within the nonrelativistic approach, the nuclear recoil effect on the energy
levels and on the wave functions of the system is accounted for by introducing two
additional terms in the Hamiltonian, traditionally referred to as the normal mass
shift (NMS) and the specific mass shift (SMS). They are given by
\begin{align}\label{4eq3}
    H_{NMS} &\,= \sum_{i} \frac{\bfp_i^2}{2M}\,,
     \\
    H_{SMS} &\,= \sum_{i<j} \frac{\bfp_i\cdot\bfp_j}{M}\,,
\end{align}
respectively, where $M$ is the mass of the nucleus. Alterations of the wave
function due to the additions to the Hamiltonian give rise to the corresponding
corrections to the hfs. The NMS part of the recoil can be factorized out and
expressed in terms of the reduced mass. It is accounted for by multiplying the
nonrelativistic hfs value by a factor of $(1+m/M)^{-3}$ \cite{bethesalpeter}. The
inclusion of $H_{NMS}$ into the CI Hamiltonian leads to the same effect but adds
some relativistic corrections. The SMS part of the recoil effect is to be
evaluated numerically, by incorporating $H_{SMS}$ into the CI Hamiltonian and by
identifying the corresponding alteration of the hfs splitting.

It should  be stressed that, despite the fact that our original CI Hamiltonian is
the relativistic one, the inclusion of the operators $H_{NMS}$ and $H_{SMS}$ in it
does not fully account for the relativistic recoil effects, since the operators
themselves are obtained within the nonrelativistic approximation only. This fact
was often disregarded in the past, e.g., in Ref.~\cite{parpia:92:recoil}. The
complete treatment of the leading [$\sim$$(\Za)^2$] relativistic recoil correction
to energy levels of the system is achieved by employing the operator
\cite{shabaev:98:rectheo}
\begin{equation}\label{4eq3a}
    H_{rec} = \frac1{2M}\sum_{ij}\left\{ \bfp_i\cdot\bfp_j -\frac{\Za}{r_i}
      \bigl[ \balpha_i+ (\balpha_i\cdot\hr_i)\hr_i\bigr]\cdot\bfp_j\right\}\,.
\end{equation}
Numerical calculations with this operator were performed, e.g., in
Ref.~\cite{tupitsyn:03}. In our present investigation, the relativistic recoil
effects are negligible as compared to other sources of the theoretical
uncertainty. We thus use the nonrelativistic operators for the description of the
recoil effects.

Matrix elements of $H_{SMS}$ between the individual CSFs can be expressed in terms
of the angular coefficients $v_{rs}^{(k)}$ introduced in Eq.~(\ref{eq7}), with the
multipolarity $k=1$,
\begin{align}\label{4eq4}
    \lbr \gamma_rPJM| H_{SMS}|\gamma_sPJM\rbr &\,= -\frac1M
  \nonumber \\ &\times
    \sum_{abcd}
   v_{rs}^{(1)}(abcd)\, V(ac)\,V(bd)\,,
\end{align}
where the radial integrals are (see, e.g., Ref.~\cite{parpia:92:recoil})
\begin{align}\label{4eq5}
    V(ac) &\,= \lbr \kappa_a||{\bm C}^{(1)}||\kappa_c\rbr\,
      \int_0^{\infty}dr\,
      \nonumber \\ & \times
      \Biggl\{ G_a \left[ \frac{d}{dr}
                -\frac{\kappa_a(\kappa_a+1)-\kappa_c(\kappa_c+1)}{2r}\right]\,G_c
      \nonumber \\ &
    + F_a \left[ \frac{d}{dr}
                -\frac{\kappa_a(\kappa_a-1)-\kappa_c(\kappa_c-1)}{2r}\right]\,F_c
                \Biggr\}\,.
\end{align}

So far, we discussed the recoil corrections to hfs that are induced by the wave
functions. There are, however, also recoil corrections to the hyperfine
interaction itself. The recoil correction to the magnetic dipole hfs interaction
arises through the spin-orbit coupling in the scalar component of the nuclear
current. This correction depends on the spin of the nucleus $I$. In the case of
hydrogen ($I=1/2$), it was derived many years ago in Ref.~\cite{barker:55},
whereas in Ref.~\cite{brodsky:67} it was reported for the case of deuterium
($I=1$). To the best of our knowledge, this correction was previously unknown  for
the arbitrary spin of the nucleus and was not accounted for in calculations of the
hfs of systems other than hydrogen and deuterium.

The spin-orbital (SO) recoil correction to the magnetic dipole hfs interaction is
obtained in Appendix by using the expression for the current of a particle with an
arbitrary spin derived in Ref.~\cite{khriplovich:96}. The result is represented by
Eq.~(\ref{aeq6}). It can be conveniently split into the normal (SON) and specific
(SOS) parts, analogously to the normal and the specific mass shift of energy
levels, $H_{SO} =  H_{SON} +  H_{SOS}$, with
\begin{align}\label{4eq6b}
    &  H_{SON} =
    \frac{\Za}{2M^2}\,(g-1)\,\bm{I}\cdot\sum_i\frac{\bm{l}_i}{r_i^3}\,, \\
        \label{4eq6bb}
    & H_{SOS} = \frac{\Za}{2M^2}\,(g-1)\,\bm{I}\cdot\sum_{i<j}
       \bm{t}_{SOS}(i,j)\,, \\
   & \bm{t}_{SOS}(i,j) = \frac{\bfr_i\times \bfp_j}{r_i^3}+\frac{\bfr_j\times \bfp_i}{r_j^3}
         \,.
\end{align}
Here, $\bm{I}$ is the operator of the nuclear spin, $\bm{l}_i$ is the operator of
the orbital angular momentum of $i$th electron, and $g$ is the g-factor of the
nucleus,
\begin{equation}\label{4eq6z}
    g = \frac{\mu}{\mu_N}\frac{M}{m_p}\frac1I\,.
\end{equation}

The SON interaction is proportional to the orbital part of the nonrelativistic
decomposition of the magnetic dipole hfs operator. It is easy to see that this
part of the SO recoil effect can be accounted for by modifying the orbital
hyperfine parameter $a_l$ by
\begin{equation}\label{4eq7a}
a_l\to a_l\left[1+\frac{m}{M}\,Z\,\frac{g-1}{g}\right]\,.
\end{equation}

It is interesting to note that, comparing to the NMS effect, the SON correction is
enhanced by a factor of $Z$, which makes it a dominant recoil effect in the hfs
splitting of medium-$Z$ H-like ions (for electronic states with $l>0$). For
Li$^{2+}$, the ratio of the SON and the NMS effects is $-0.5$ for the $2p_{1/2}$
state and $-1.2$ for the $2p_{3/2}$ state. For the lithium-like systems, however,
the SON and the SOS corrections tend to cancel each other, the net effect being
rather small numerically.

In the present work, we calculate the SOS correction by using perturbation theory
to the lowest order. For the electronic configuration with a single valence
electron beyond the closed core shell, the contribution to the function $G_{M1}$
due to the SOS effect can be expressed as
\begin{align}\label{4eq7b}
  \delta G_{M1}^{SOS}  = &\ \left[ \frac{(\Za)^3m^3}{n^3J(J+1)(2L+1)}
  \right]^{-1}\,
    \nonumber \\ &
       \times
     \frac{Zm}{M}\,\frac{g-1}{g}\,
       \sum_{\mu_c} (-1)\lbr cv| t_{{SOS}_0}|vc\rbr\,,
\end{align}
where $v$ denotes the valence electron state with the angular momentum projection
$\mu_v = 1/2$, $c$ is the core electron state with the angular momentum projection
$\mu_c$, and $t_{{SOS}_0}$ is the zeroth spherical component of the operator
$\bm{t}_{SOS}$. The radial integral is evaluated to yield
\begin{align}\label{4eq7c}
\lbr cv| t_{{SOS}_0}|vc\rbr &\ =
  \sqrt{\frac{6}{j_v(j_v+1)(2j_v+1)}} \,
   \nonumber \\ &\times
    \SixJ{1}{1}{1}{j_v}{j_v}{j_c}\, U(cv)\, V(vc)\,,
\end{align}
where
\begin{align}\label{4eq7d}
U(cv) = \lbr \kappa_c||\bm{C}^{(1)}||\kappa_v\rbr\,\int_0^{\infty}
   dr\, r^{-2}\,\bigl( G_cG_v+F_cF_v\bigr)\,,
\end{align}
and $V(vc)$ is defined by Eq.~(\ref{4eq5}). It is easy to see that the radial
integral (\ref{4eq7c}) vanishes for the $S$ states.

The calculational results for the individual recoil contributions to the magnetic
dipole hfs splitting are listed in Table~\ref{tab:dipole} under entries ``NMS'',
``SMS'', and ``SO''. The results obtained for the NMS and SMS parts are in good
agreement with the previous evaluations of these corrections. The entry ``SO''
represents the sum of the SON and the SOS corrections. Because of a large
cancellation between these two parts, we calculate both of them by perturbation
theory. The uncertainty specified in the table was evaluated by comparing results
obtained with different potentials in the zeroth-order Hamiltonian.

The scalar component of the nuclear current yields also a correction to the
electric quadrupole interaction \cite{brodsky:67,khriplovich:96}. This correction
is induced by the nuclear spin and can be interpreted as a shift of the nuclear
quadrupole moment (see Appendix for details). The induced contribution is included
into the observable value of the nuclear quadrupole moment and thus is not needed
to be taken into account in the theoretical description of the electric quadrupole
hfs.

\subsection{Nuclear size and magnetization distribution}

Due to a high singularity of the hfs interaction at the origin, the nuclear
structure effects (particularly, the distribution of the nuclear magnetic moment)
have significant influence on the magnetic dipole hfs and should be taken into
account in atomic calculations. An accurate theoretical description of these
effects is a demanding problem. A way for its rigorous solution was paved in
recent studies \cite{friar:05:prc,friar:05:plb,pachucki:07:nphfs}. Practical
realizations of this approach, however, are so far restricted to two- and
three-nucleon systems \cite{friar:05:prc,friar:05:plb} and their extension to more
complex nuclei like $^7$Li and $^9$Be looks problematic.

The most widely used approach up to now is to account for the extended nuclear
magnetization distribution [the Bohr-Weisskopf (BW) effect] by means of the Zemach
formula \cite{zemach:56}. According to the original formulation, the nuclear
correction to the magnetic dipole hfs of an $S$ state of an H-like atom is
represented by a simple multiplicative factor,
\begin{equation}\label{5eq1}
  \delta G_{M1}^{nuc}(Z) = -2\Za \, \lbr r\rbr_{em}\,G_{M1}(Z)\,,
\end{equation}
where $\lbr r\rbr_{em}$ is the Zemach moment obtained by folding together the
electric charge $\rho_e(r)$ and magnetization $\rho_m(r)$ densities
\begin{equation}\label{5eq2}
   \lbr r\rbr_{em} = \int d\bfr d\bfr\pr\,
       \rho_e(r)\,\rho_m(r\pr) \,|\bfr-\bfr\pr|\,.
\end{equation}

Formula (\ref{5eq1}) accounts for both the charge and the magnetization
distribution. Since the charge distribution effect is usually taken into account
in a more complete way by modifying the Coulomb nuclear potential in the
Hamiltonian (\ref{eq2}), it should be subtracted from the total Zemach correction.
The finite nuclear charge (FNC) correction is obtained from Eq.~(\ref{5eq1}) by
setting $\lbr r\rbr_{em} = \lbr r\rbr_{e}$, where $\lbr r\rbr_{e}$ is the electric
charge radius defined as
\begin{equation}\label{5eq6}
    \lbr r\rbr_{e} = \int d\bfr
       \rho_e(r)\, \,|\bfr|\,.
\end{equation}
More detailed studies of the FNC correction in H-like atoms with including the
relativistic effects were reported in Refs.~\cite{shabaev:94:hfs,volotka:03}. For
lithium and beryllium, the relativistic effects are small and enter mainly through
the alteration of the exponent in the $\Za$ and $\lbr r\rbr_{e}$ dependence  by
terms $\sim(\Za)^2$.

Using the hydrogenic result \cite{shabaev:94:hfs} for the exponent of the $\Za$
and $\lbr r\rbr_{e}$ dependence, we write the generalization of the
nonrelativistic FNC correction in the form valid for an arbitrary state of
few-electron atoms
\begin{equation}\label{5eq7}
    \delta G_{M1}^{FNC}(Z) = -2\, (\Za\, \lbr r\rbr_{e})^{2\gamma-1}\,G_{M1,c}(Z)\,,
\end{equation}
where $\gamma = \sqrt{1-(\Za)^2}$ and $G_{M1,c}$ is the contact part of $G_{M1}$.
Using Eq.~(\ref{5eq7}), one should keep in mind that the charge radius $\lbr
r\rbr_{e}$ is different from the charge root-mean-square (rms) radius $\rms_e$,
which is usually listed in tables. The conversion factor depends somewhat on the
model of the nuclear charge distribution. For the Gaussian model, the connection
is $\lbr r\rbr_{e} = \sqrt{8/(3\pi)}\,\rms_e$. The values of the rms radii of the
nuclei $^{6,7}$Li and $^9$Be were taken from Ref.~\cite{angeli:04}. The rms radii
and the numerical results for the FNS correction obtained by Eq.~(\ref{5eq7}) are
presented in Table~\ref{tab:nucl} (the relative values of the correction, $\delta
G_{M1}^{FNC}/G_{M1,c}$, are listed). An independent evaluation of the FNS
correction was performed, by repeating the CI calculations for different values of
the nuclear radius and by fitting the increments to the analytical form
(\ref{5eq7}). The FNS correction obtained in this way agreed very well with the
analytical results presented in the table.

In the present investigation, separate values of the FNS correction are not
necessary since this effect is already included in the CI part of the calculation.
However, we use the values of the FNS correction in order to identify the
point-nucleus limit of our results (particularly, for the comparison with the
point-nucleus results of Ref.~\cite{yan:96}) and for improving the stability of
the fit in extracting the nonrelativistic limit of our calculations (the FNC
correction is the only part of the CI values that is linear in $\alpha$).

There is no need to specify explicitly to which electronic states the results for
the nuclear corrections in Table~\ref{tab:nucl} correspond, because the relative
values of the corrections are listed. Our numerical calculations show that, with a
good accuracy, the relative values of the nuclear corrections do not depend on the
particular state. (Of course, for the $P$ states, the relative value should be
evaluated with respect to the contact part of the correction.)

\begin{table*}
\setlength{\LTcapwidth}{\textwidth}
\caption{Nuclear parameters (in fm) and the relative values of  the nuclear
corrections (in ppm) to the magnetic dipole hfs. The abbreviations are as follows:
``FNS'' denotes the nuclear charge distribution correction, ``Zemach'' labels the
BW correction obtained with the Zemach formula, ``SP'' labels the BW correction
evaluated within the SP approach, ``const'' indicates that the odd-nucleon wave
function is taken to be a constant within the nucleus. \label{tab:nucl}}
\begin{ruledtabular}
\begin{tabular}{c.........}
Isotope
 &\multicolumn{1}{c}{$\rms_e$}            &\multicolumn{1}{c}{$\rms_m$}           &\multicolumn{1}{c}{$\lbr r\rbr_{em}-\lbr r\rbr_{e}$}
       &\multicolumn{1}{c}{FNS} & \multicolumn{4}{c}{Bohr-Weisskopf effect} &  \multicolumn{1}{c}{Total nuclear} \\
 &\multicolumn{1}{c}{[fm]} & \multicolumn{1}{c}{[fm]} & \multicolumn{1}{c}{[fm]} &
        & \multicolumn{1}{c}{Zemach} & \multicolumn{1}{c}{SP} & \multicolumn{1}{c}{SP(const)} &  \multicolumn{1}{c}{Total} &
        \multicolumn{1}{c}{correction}\\
\hline\\[-9pt]
$^6$Li &  2.5x4(3) &  3.1x2(22) &  1.3x8(19) & -268x(3) &  -160x(20) &    -x50 &   -x89 &  -100x(60) &  -368x(60) \\
$^7$Li &  2.4x3(3) &  2.8x0(8)  &  1.1x9(8)  & -257x(3) &  -135x(9)  &   -1x12 &   -x99 &  -112x(23) &  -369x(23) \\
$^9$Be &  2.5x2(1) &  2.6x7(6)  &  1.0x6(5)  & -356x(4) &  -160x(8)  &   -1x58 &  -1x61 &  -158x(16) &  -514x(16) \\
\end{tabular}
\end{ruledtabular}
\end{table*}

The Zemach correction induced by the magnetization distribution (the BW effect)
can be written in a form valid for an arbitrary state as
\begin{equation}\label{5eq8}
  \delta G_{M1}^{BW}(Z) = -2\Za \, \left[\lbr
  r\rbr_{em}-\lbr r\rbr_{e}\right]\,G_{M1,c}(Z)\,.
\end{equation}
The Zemach radius is usually not tabulated and should be derived from data
available for the charge and magnetization rms radii according to
Eq.~(\ref{5eq2}), with an additional input of the distribution models. For the
Gaussian model
\begin{equation}\label{5eq9a}
    \rho(r) = \rho_0\, \exp (-\Lambda r^2)\,
\end{equation}
employed for the charge and magnetization distributions, the Zemach radius is
readily obtained analytically,
\begin{equation}\label{5eq9b}
   \lbr r\rbr_{em} = \sqrt{\frac{8}{3\pi}}\,\left(\rmsSq_e+\rmsSq_m\right)^{1/2}\,.
\end{equation}
For more sophisticated distribution models, one has to evaluate radial
integrations in Eq.~(\ref{5eq2}) numerically. In order to test the model
dependence of the Zemach radius (with fixed values of the charge and magnetization
rms radii), we performed its numerical evaluation with the two-parameter Fermi
model. The same results as for the Gaussian model are obtained, which leads us to
conclude that the model dependence is negligible.

The values listed in Table~\ref{tab:nucl} for the magnetic rms radius are the
average of data tabulated in Ref.~\cite{jager:74} and the errors are their
mean-square deviation. Under the entry ``Zemach'', we tabulate the numerical
results for the BW correction obtained by Eq.~(\ref{5eq8}); the error ascribed to
them originates from the uncertainties of the magnetization and charge radii.

The second approach to the description of the BW effect considered in the present
work is based on the single particle (SP) model of the nuclear magnetic moment and
will be referred to as the SP approach in the following. Within the SP model, the
nuclear magnetic moment is assumed to be induced by the odd nucleon (proton, when
$Z$ and $A$ are odd and neutron, when $Z$ is even and $A$ is odd). The odd nucleon
is assumed to have an effective $g$ factor, which is fixed so that it yields the
experimental value of the nuclear magnetic moment. The treatment of the
magnetization distribution effect on hfs within the SP model was originally
developed in Refs.~\cite{bohr:50,bohr:51} and later in Ref.~\cite{shabaev:94:hfs}.
The spin-orbit interaction of the odd nucleon was introduced into this approach in
Ref.~\cite{shabaev:97:pra}. Our present treatment closely follows the procedure
described in Refs.~\cite{shabaev:97:pra,zherebtsov:00:cjp}.

The wave function of the odd nucleon is assumed to satisfy the Schr\"odinger
equation with the central potential of the Woods-Saxon form and the spin-orbital
term included (see, e.g., Ref.~\cite{elton:67})
\begin{equation}\label{5eq10}
    V(\bfr) = -V_0\,{\cal F}(r)+ \frac1{m_p}\phi_{so}(r)\,\bm{l}\cdot\bsigma+ V_C(r)\,,
\end{equation}
where
\begin{equation}\label{5eq10a}
    \phi_{so}(r) = \frac{V_{so}}{4m_pr}\,
    \frac{d{\cal F}(r)}{dr}\,,
\end{equation}
\begin{equation}\label{5eq10b}
    {\cal F}(r) = \left[1+\exp\left(\frac{r-R}{a} \right)\right]^{-1}\,,
\end{equation}
and $V_C$ is the Coulomb part of the interaction (absent for neutron), with the
uniform distribution of the charge ($Z-1$) over the nuclear sphere. The parameters
$V_0$, $V_{so}$, $R$, and $a$ were taken from Ref.~\cite{elton:67}, where they
were obtained by fitting electron scattering data. The fitting was not evaluated
for $^9{\rm Be}$, so we use the parameters for its closest odd-neutron neighbour,
$^{12}{\rm C}$. For $^6{\rm Li}$, the nuclear spin is integer ($I=1$), and so one
needs to make an additional assumption about the value of the orbital angular
momentum of the nucleus. We used the value $L=0$ \cite{nemirovsky:book} in our
calculations.

The nuclear magnetic moment can be evaluated within the SP model to yield
\cite{shabaev:97:pra}
\begin{align}\label{5eq11}
    \frac{\mu}{\mu_N} =
     \begin{aligned}
      \displaystyle
       \frac12\, g_S+ \left[I-\frac12+\frac{2I+1}{4(I+1)}\lbr\phi_{so}r^2\rbr
       \right] g_L\,, \ \ \ \ \ \ \ \ \ \ \ & \\
        \ \mbox{for} \  \ I = L+\frac12\,, & \\
      \displaystyle
       -\frac{I}{2(I+1)}\, g_S+ \left[\frac{I(2I+3)}{2(I+1)}
        -\frac{2I+1}{4(I+1)}\lbr\phi_{so}r^2\rbr
       \right]& \ g_L\,,\ \\
       \ \mbox{for} \  \ I = L-\frac12\,, & \\
     \end{aligned}
\end{align}
where $I$ and $L$ are the total and the orbital angular momentum of the nucleus,
respectively, $g_L$ is the $g$ factor associated with the orbital motion of the
nucleon ($g_L$ = 1 for proton and $g_L = 0$ for neutron) and $g_S$ is the
effective nucleon $g$ factor, determined by the condition that Eq.~(\ref{5eq11})
yields the experimental value of the magnetic moment.

It was demonstrated in Ref.~\cite{shabaev:97:pra} that, within the SP model, the
BW effect can be accounted for by adding a multiplicative
magnetization-distribution function $F(r)$ to the standard point-dipole hfs
interaction (\ref{2eq2}). The distribution function is given by
\cite{zherebtsov:00:cjp}
\begin{widetext}
\begin{eqnarray} \label{5eq12}
  F(r) &=& \frac{\mu_N}{\mu}
    \int_0^{r}dr\pr\, {r\pr}^2 |u(r\pr)|^2\,
         \left[ \frac12\, g_S+ \left(I-\frac12+\frac{2I+1}{4(I+1)}\,r^2\phi_{so}(r)
       \right) g_L \right]
  \nonumber \\ &&
  + \frac{\mu_N}{\mu} \int_r^{\infty}dr\pr\, {r\pr}^2 |u(r\pr)|^2\,\frac{{r}^3}{{r\pr}^3}\,
         \left[ -\frac{2I-1}{8(I+1)}\, g_S+ \left(I-\frac12+\frac{2I+1}{4(I+1)}\,r^2\phi_{so}(r)
       \right) g_L \right]\,,
\end{eqnarray}
for $I = L+1/2$ and
\begin{eqnarray} \label{5eq13}
  F(r) &=& \frac{\mu_N}{\mu}
    \int_0^{r}dr\pr\, {r\pr}^2 |u(r\pr)|^2\,
         \left[ -\frac{I}{2(I+1)}\, g_S+ \left(\frac{I(2I+3)}{2(I+1)}-\frac{2I+1}{4(I+1)}\,r^2\phi_{so}(r)
       \right) g_L \right]
  \nonumber \\ &&
  + \frac{\mu_N}{\mu} \int_r^{\infty}dr\pr\, {r\pr}^2 |u(r\pr)|^2\,\frac{{r}^3}{{r\pr}^3}\,
         \left[ \frac{2I+3}{8(I+1)}\, g_S+ \left(\frac{I(2I+3)}{2(I+1)}-\frac{2I+1}{4(I+1)}\,r^2\phi_{so}(r)
       \right) g_L \right]\,,
\end{eqnarray}
\end{widetext}
for $I = L-1/2$. In the above formulas, $u(r)$ is the wave function of the odd
nucleon. It can easily be seen that $F(r) = 1$ outside the nucleus.

In the present work, we evaluated the BW correction within the SP approach as
described above. In addition, we considered a simplified version of this approach
obtained by assuming the wave function of the odd nucleon to be just a constant
within the nucleus. By comparing the two corresponding results, we can
conservatively estimate the dependence of the SP values on the parameters employed
in the Woods-Saxon potential. Calculational results for the BW correction obtained
within the SP approach are listed in Table~\ref{tab:nucl} under the labels ``SP"
(the full SP approach) and ``SP(const)'' (the SP approach with the constant wave
function of the odd nucleon).

The total results for the BW correction listed in the table were obtained as
follows. For $^7$Li and $^9$Be, we employ the results of the SP model as the final
values. The uncertainty for $^7$Li was taken to be the largest deviation from the
final result. For $^9$Be, all three values fall very close to each other, so we
assign the 10\% uncertainty to the final result. The nucleus $^6$Li has an odd
neutron and an odd proton; one thus can hardly expect it to be described well by
the SP model. In this case, we use the plain average of the three values; the
error was chosen so that it covers all three results.

We would like to stress that there are nontrivial nuclear structure effects, which
are ignored both within the SP model and within the Zemach approach. Since these
effects cannot be reasonably estimated at present, our uncertainties of the BW
correction yield the order of the expected error only.

The final values of the BW correction and their uncertainties are listed in
Table~\ref{tab:dipole} under the entry ``BW''. Our results for $^7$Li and $^9$Be
are reasonably close to the Zemach-formula values of Ref.~\cite{yan:96}. The
result of Ref.~\cite{bieron:99} for $^9$Be is larger than ours by a factor of 4.
This is because the authors of Ref.~\cite{bieron:99} used $\lbr r\rbr_m$ instead
of $\lbr r\rbr_{em}-\lbr r\rbr_{e}$ in Eq.~(\ref{5eq8}). In Ref.~\cite{king:07},
the BW correction was evaluated within the SP approach with the constant
odd-nucleon wave function. The corresponding results for $^{6,7}$Li nearly
coincide with our values obtained within the same approach. In
Refs.~\cite{boucard:00,sapirstein:03:hfs}, the BW correction was calculated by
using the Fermi \cite{boucard:00} or the uniform \cite{sapirstein:03:hfs}
distribution of the magnetization density over the nucleus. Their results fall
between our values obtained with different models. In most of other previous
studies, the BW effect was not accounted for.

\subsection{Negative continuum}

The negative-continuum (NC) contribution might be of some importance in
calculations involving the operators that mix the upper and the lower components
of the Dirac wave function. The magnetic dipole hfs operator is of this kind, so
we have to obtain an estimation for this correction. In the present investigation,
we calculate the NC contribution by employing perturbation theory to the first
order.

For the electronic configuration with a single valence electron beyond the closed
core shell and to first order in the electron-electron interaction, the NC
correction can be written as
\begin{align} \label{6eq1}
  \delta G_{M1}^{NC} &\ = \left[ \frac{(\Za)^3m^2}{n^3J(J+1)(2L+1)}
  \right]^{-1}\, 2 \sum_{n}^{\vare_n < 0}
    \nonumber \\ &
  \times \Biggl\{
    \sum_{\mu_c} \frac{\left[ \lbr vc|V_{CB}|nc\rbr - \lbr cv|V_{CB}|nc\rbr \right] \lbr n|{{t}}^{(1)}_0|v\rbr }
         {\vare_v-\vare_n}
    \nonumber \\ &
    +     \sum_{\mu_c} \frac{\left[ \lbr vc|V_{CB}|vn\rbr - \lbr cv|V_{CB}|vn\rbr \right] \lbr n|{{t}}^{(1)}_0|c\rbr }
         {\vare_c-\vare_n}
    \nonumber \\ &
    - \frac{\lbr v|U|n\rbr\, \lbr n| {\bm{t}}^{(1)}_0|v\rbr }{\vare_v-\vare_n}
     \Biggr\}\,,
\end{align}
where $v$ denotes the valence electron state with the angular momentum projection
$\mu_v = 1/2$, $c$ is the core electron state with the angular momentum projection
$\mu_c$, ${{t}}^{(1)}_0$ is the spherical component of the magnetic dipole hfs
operator ${\bm{t}}^{(1)}$ defined by Eq.~(\ref{2eq2}), $V_{CB} = V_C+V_B$ is the
sum of the Coulomb and Breit parts of the electron-electron interaction, and the
summation over $n$ is performed over the negative-energy part of the Dirac
spectrum. The states $v$, $c$, and $n$ are assumed to be eigenvectors of a
single-particle Dirac Hamiltonian $h$ with the screening potential $U$, $h =
\balpha\cdot\bfp+ (\beta-1)m+ V_{\rm nuc}(r)+ U(\bfr)$.

We mention that the NC correction may depend strongly on the choice of the
Hamiltonian $h$. It is, therefore, important to use the same single-particle
Hamiltonian for the evaluation of the NC correction as in the CI part of the
calculation. So, in the present investigation the screening potential $U$ in
Eq.~(\ref{6eq1}) was fixed as $U = V^{N-1}_{DF}$.

It should be also noted that, evaluating Eq.~(\ref{6eq1}), one cannot neglect the
Breit part of the electron-electron interaction as compared to the Coulomb one.
For the negative-energy part of the Dirac spectrum, both of these interactions
induce contributions of the same order of magnitude.

Formula (\ref{6eq1}) for the NC correction ignores contributions of the second and
higher orders in the electron-electron interaction. Their unambiguous description
is possible within QED only. For lithium and beryllium, the electron correlation
is strong and the perturbation expansion converges slowly. We thus assign the
uncertainty of 50\% to the NC contribution obtained by Eq.~(\ref{6eq1}).

\section{Discussion}
\label{sec:discussion}

The calculational results for the magnetic-dipole hfs splitting of the $2^2S$,
$3^2S$, $2^2P_{1/2}$, and $2^2P_{3/2}$ states of $^{6,7}$Li and $^9$Be$^+$ are
listed in Table~\ref{tab:dipole}, expressed in terms of the dimensionless function
$G_{M1}$ defined by Eq.~(\ref{2eq7}). The values presented for the $S$ states
differ from those in our previous work \cite{yerokhin:08:pra} in two ways. First,
we now treat the normal mass shift as a correction, rather than by including it
into the definition of the function $G_{M1}$. (Of course, this difference has no
effect on the total theoretical prediction for the hfs or the hyperfine constant
$A_J$.) Second, we perform a more detailed analysis of the BW effect, and so the
uncertainty of this correction is changed.

In order to convert the function $G_{M1}$ into the hyperfine constant $A_J$, an
additional experimental input in the form of the magnetic moment of the nucleus is
needed. This issue might contain some ambiguities since the tabulated values of
the nuclear magnetic moments \cite{stone:05} are often inconsistent. In the case
of $^6{\rm Li}$ and $^7{\rm Li}$, we assume the values originally obtained by
Beckmann {\em et al.} \cite{beckmann:74} to be the most reliable ones. For $^9$Be,
the choice is less obvious, and we present the theoretical results for the
hyperfine constant $A_J$ for two different experimental values of the magnetic
moment.

For $^7$Li, we observe good agreement of our theoretical predictions with all the
experimental results listed, except the recent measurement of the $2^2P_{1/2}$
state by Das and Natarajan \cite{das:08}, which claims to be accurate to better
than 0.01\%. The theoretical prediction is away from this measurement by about
18$\sigma$ and we presently see no way to explain this deviation theoretically.

Our theoretical prediction for the ground-state hfs splitting of $^6$Li is in
slight disagreement with the high-precision experimental result. For the ground
state of beryllium, the deviation is larger and amounts to 3 or 4$\,\sigma$,
depending on the value of the nuclear magnetic moment used. There are two possible
explanations of these discrepancies: underestimated systematic effects in the
experimental values of the nuclear magnetic moments and nontrivial
nuclear-structure effects in the theoretical predictions. Basing on the
experimental data available, we cannot unambiguously distinguish between these two
explanations. It would have been possible if the hfs splitting of two different
states were accurately measured for the same isotope. The existing measurements of
the hfs of excited states, however, are not yet sensitive to the inconsistencies
in values of the nuclear magnetic moment.

The comparison presented for the hfs of excited states of lithium and beryllium
indicates that our theoretical predictions are more accurate than the experimental
results. The general agreement with the experimental data for the excited states
is good, the only exception being the results of Ref.~\cite{das:08}.

Our calculational results for the electric quadrupole hfs splitting of the
$2^2P_{3/2}$ state in $^{6,7}$Li and $^9$Be$^+$ are listed in
Table~\ref{tab:quadrupole}, expressed in terms of the dimensionless function
$G_{E2}$ defined by Eq.~(\ref{3eq15}). It is remarkable that all theoretical
contributions to the electric quadrupole splitting  seem to be well under control,
so that the resulting theoretical predictions for the function $G_{E2}$ are
obtained with very good accuracy. If accurate experimental investigations of the
quadrupole splitting were possible for Li-like ions, they would lead to a
high-precision determination of the nuclear quadrupole moments, which are
difficult to measure directly. In the absence of such investigations, the
theoretical predictions for the hyperfine constant $B_J$ are obtained by using the
tabulated values of the nuclear quadrupole moments \cite{stone:05}. Our results
are in agreement with the scarce experimental data available.

\section*{Acknowledgements}

I wish to thank V.~M.~Shabaev for valuable advices and, particularly, for pointing
out the presence of the two-electron part in the spin-orbital recoil correction.
Helpful discussions with K.~Pachucki  and A.~Derevianko are gratefully
acknowledged. The work presented in this paper was supported by the ``Dynasty''
foundation and by RFBR (grant No.~06-02-04007).

\appendix*

\section{Corrections to the hyperfine splitting due to the scalar component
 of the nuclear current}

Let us first consider the simplest case of the electron and the nucleus being the
spin-$1/2$ Dirac particles. The electron-nucleus interaction is then given by the
standard scattering amplitude \cite{LL4}
\begin{equation}\label{aeq0}
   M = -e^2Z\, D_{\mu\nu}(q)\, j_e^{\mu}\,j_{nuc}^{\nu}\,,
\end{equation}
where $D^{\mu\nu}$ is the photon propagator and $j_e$ and $j_{nuc}$ are the
electromagnetic current of the electron and the nucleus, respectively,
\begin{align}\label{aeq1}
    j^{\mu} &\,= \overline{u}(p\pr)\, \gamma^{\mu} \,u(p)
  \nonumber \\ &
     = \frac1{2m} \overline{u}(p\pr)\,\left[ (p\pr+p)^{\mu}
        + i \sigma^{\mu\nu}q_{\nu}\right]\,u(p)\,.
\end{align}
Here, $u$ is the free Dirac spinor, $m$ is the mass of the particle, $q = p\pr-p$,
and $\sigma^{\mu\nu} = (i/2)[\gamma_{\mu},\gamma_{\nu}]$. Expressing the time
component of the nuclear current in terms of the free spinors $w$ in the rest
frame, one arrives at \cite{LL4}
\begin{equation}\label{aeq2}
    j^0_{nuc} = w^*\, \left(1-\frac{q^2}{8M^2}+ \frac{i\bsigma \cdot \bfq\times
    \bfp}{4M^2}\right)\,w\,,
\end{equation}
where $M$ is the nuclear mass. The first term in the brackets in the above
expression corresponds to the standard Coulomb interaction between electron and
the nucleus. The third term represents the spin-orbital coupling and induces a
recoil correction to the hfs we are interested here.

The generalization of the expression (\ref{aeq1}) for the case of the nucleus with
an arbitrary spin was obtained in Ref.~\cite{khriplovich:96}. The time component
of the current reads
\begin{equation}\label{aeq3}
j^0_{nuc} = \frac1{2M}\, \overline{\psi}(p\pr)\, \left[F_e\,(E+E\pr)  +
G_m\,\gamma^0
  \bm{\Gamma}\cdot\bfq\, \right] \, \psi(p)\,,
\end{equation}
where $F_e$ and $G_m$ are the electric and magnetic form factors of the nucleus,
respectively,
\begin{equation}\label{aeq4}
    \bm{\Gamma} = \left(
        \begin{array}{cc}
        0  & \bm{\Sigma}  \\
       -\bm{\Sigma}  & 0   \\
        \end{array}
     \right)\,,
\end{equation}
and the vector $\bm{\Sigma}$ is constructed from components $\Sigma_i$, which are
generalizations of the Pauli matrices. After expressing the current in terms of
the spinors $\xi_0$ in the rest frame, one gets the generalization of
Eq.~(\ref{aeq2}) to the case of an arbitrary-spin nucleus \cite{khriplovich:96},
\begin{align} \label{aeq5}
j^0_{nuc} = &\, {\xi_0\pr}^*\,\left[
 F_e - (2G_m-F_e)\,\frac{(\bm{\Sigma}\cdot\bfq)^2}{8M^2}
       \right.
  \nonumber \\ &
        \left.
   + (2G_m-F_e)\,i \frac{\bm{I}\cdot (\bfq\times \bfp)}{2M^2}
 \right] \, \xi_0\,,
\end{align}
where $I$ is the operator of the spin of the nucleus. The form factors are
identified as \cite{khriplovich:96}: $F_e(0) = 1$ and $G_m(0) = g/2$, where $g$ is
the $g$ factor of the nucleus.

The third term in the brackets of Eq.~(\ref{aeq5}) induces a first-order (in the
electron-nucleus mass ratio) recoil correction to the magnetic dipole hfs
splitting. Taking into account that, in the center-of-mass system, the total
momentum of the atom is zero and transforming this term into the coordinate space
(see Ref.~\cite{LL4}), we obtain the interaction of the form
\begin{equation}\label{aeq6}
    H_{SO} = \frac{\Za}{2M^2}\,(g-1)\,\bm{I}\cdot
        \sum_i  \Biggl( \frac{\bfr_i \times \bfp_i}{r_i^3}
          + \sum_{j\ne i} \frac{\bfr_i \times \bfp_j}{r_i^3}
             \Biggr)\,,
\end{equation}
where indices $i$ and $j$ numerate the electrons in the atom. For the hydrogen
atom, $\bm{I} = (1/2)\bsigma$, Eq.~(\ref{aeq6}) reproduces the well-known result
of Ref.~\cite{barker:55}. A similar recoil correction to the Zeeman splitting of
multi-electron atoms was reported in Ref.~\cite{phillips:49}.

The second term in the brackets of Eq.~(\ref{aeq5}) can be split into the contact
and the quadrupole part, which induce corrections to the Lamb shift and to the
quadrupole hfs splitting, respectively. Both of these corrections were evaluated
in Ref.~\cite{khriplovich:96}. The result for the quadrupole interaction due to
the second term in Eq.~(\ref{aeq5}) is (with the additional factor of 2, corrected
in Ref.~\cite{pomeranskii:98})
\begin{equation}\label{aeq7}
    \delta H_{E2} = \frac{\Za(g-1)}{2M^2}\,\Lambda\,\nabla_i\nabla_j \frac1r
       \left( I_iI_j-\frac13\delta_{ij}I^2\right)\,,
\end{equation}
where
\begin{equation}\label{aeq7a}
    \Lambda = \begin{array}{lc}
           \displaystyle
                 1/(2I-1)\,, & I \ \ \mbox{is integer}\,, \\
           \displaystyle
                 1/(2I)\,, & I \ \ \mbox{is half-integer}\,.
              \end{array}
\end{equation}
Using the standard angular-momentum algebra, we transform Eq.~(\ref{aeq7}) into
the form analogous to Eq.~(\ref{3eq4}),
\begin{equation}\label{aeq8}
 \delta H_{E2} = \alpha \,\bm{T}^{(2)}\cdot \delta \bm{Q}^{(2)}\,,
\end{equation}
where $ \delta \bm{Q}^{(2)}$ is the correction to the operator of the quadrupole
moment,
\begin{equation}\label{aeq9}
     \delta \bm{Q}^{(2)} = -\frac{Z(g-1)\Lambda}{2M^2}\,\sqrt{6}\,(I\otimes
     I)^{(2)}\,.
\end{equation}
Taking into account Eq.~(\ref{3eq12}), the correction to the nuclear quadrupole
moment is identified, which is
\begin{equation}\label{aeq10}
    \delta Q =\begin{array}{lc}
           \displaystyle
                 -\frac{Z(g-1)I}{M^2}\,, & I \ \ \mbox{is integer}\,, \\ \\
           \displaystyle
                 -\frac{Z(g-1)(I-1/2)}{M^2}\,, & I \ \ \mbox{is half-integer}\,.
              \end{array}
\end{equation}

The numerical values of the induced quadrupole moment for the isotopes considered
in this work are: $\delta Q_{\rm ind}(^7{\rm Li}) = -0.39$~mbarn, $\delta Q_{\rm
ind}(^6{\rm Li}) = -0.15$~mbarn, and $\delta Q_{\rm ind}(^9{\rm Be}) =
-0.18$~mbarn, to be compared with the total values of the nuclear quadrupole
moments \cite{stone:05}: $Q(^7{\rm Li}) = -40.55\,(80)$~mbarn, $Q(^6{\rm Li}) =
-0.82\,(2)$~mbarn, $Q(^9{\rm Be}) = 52.88\,(38)$~mbarn.

It should be mentioned that the correction (\ref{aeq10}) does not have immediate
experimental consequences. It vanishes for the nuclear spin $I=0$ and $1/2$.
Nuclei with the spin $I>1/2$ have a quadrupole moment, and so the correction
(\ref{aeq10}) appears only together with the ``pure'' nuclear quadrupole moment.
If the values of the nuclear quadrupole moments are derived from experimental
observations, the induced correction is included in them and thus does not have to
be accounted for in theoretical descriptions of the quadrupole splitting. It
should be included, however, when the nuclear quadrupole moments are calculated
basing on microscopic nuclear models, as, e.g., in Ref.~\cite{mertelmeier:86}.



\end{document}